\documentclass[aps,prl,showkeys,nofootinbib,amsmath,amssymb,superscriptaddress,twocolumn]{revtex4-2}
\usepackage{dsfont}
\usepackage{mathtools} 
\usepackage{xcolor}
\definecolor{nicegreen}{rgb}{0.07, 0.564, 0.04}
\usepackage{framed}
\usepackage{comment}
\usepackage{xspace}
\usepackage{graphicx}
\usepackage{dcolumn}
\usepackage{bm}
\usepackage{tikz-cd}

\newcommand\ie{\mbox{\textit{i.\,e.}}\xspace}

\newcommand{\hP}{\hat{P}}

\DeclarePairedDelimiter\braket{\langle}{\rangle}
\DeclarePairedDelimiterX\Braket[2]{\langle}{\rangle}{#1 \delimsize\vert #2}
\definecolor{nicegreen}{rgb}{0.07, 0.564, 0.04}

\begin{document}

\title{The minimal length is physical}

\author{Pasquale Bosso}\email[]{bosso.pasquale@gmail.com}
\affiliation{University of Lethbridge, Department of Physics and Astronomy, 4401 University Drive, Lethbridge, Alberta, Canada, T1K 3M4}
\author{Luciano Petruzziello}\email[]{lupetruzziello@unisa.it}
\affiliation{Dipartimento di Ingegneria Industriale, Universit\`a degli Studi di Salerno, Via Giovanni Paolo II, 132 I-84084 Fisciano (SA), Italy \\
INFN, Sezione di Napoli, Gruppo collegato di Salerno, Via Giovanni Paolo II, 132 I-84084 Fisciano (SA), Italy}
\author{Fabian Wagner}
\email[]{fabian.wagner@usz.edu.pl}
\affiliation{Institute of Physics, University of Szczecin, Wielkopolska 15, 70-451 Szczecin, Poland\\
CP3-Origins, University of Southern Denmark, Campusvej 55, DK-5230 Odense M, Denmark}
\date{\today}

\begin{abstract}
In this paper, we clarify a foundational loose end affecting the phenomenological approach to quantum gravity centered around the generalization of Heisenberg uncertainty principle. This misconception stems from a series of recently published works in which perturbative and non-perturbative methods are confused, thereby resulting in a blurring of the distinction between changes in the deformed algebra and changes in the representation of operators.
Accordingly, this reasoning would render the existence of a minimal length representation-dependent, and thus unphysical.
\end{abstract}

\maketitle


\section{Introduction}

The existence of a minimal length scale has been predicted by several promising candidates for a quantum theory of the gravitational interaction \cite{Mead:1964zz,Gross:1987kza,Gross:1987ar,Amati:1987wq,Amati:1988tn,Konishi:1989wk,Garay:1994en,Adler:1999bu,Scardigli:1999jh}.
One of the most acknowledged implications of the above phenomenon results in a phenomenological approach to quantum gravity that describes a quantum mechanical setting with a deformed Heisenberg algebra. The aforementioned treatment of this major open problem in theoretical physics is commonly referred to as generalized uncertainty principle (GUP). Since the earliest achievements obtained in the context of string theory, this phenomenological model has been widely investigated in a plethora of different scenarios, leading to considerable theoretical accomplishments and pointing towards challenging future perspectives.

Recently, it has been argued in a series of articles \cite{b3,b4,Bishop22,Bishop:2022des} that there exists a remarkable ambiguity in the GUP scheme. A blunder of this kind should, in principle, avert the identification of a minimal length scale by merely deforming the canonical commutation relations. The harbingers of a similar argument can already be found in other works \cite{b1,b2} in which, however, the critique to the GUP is missing.

In particular, it has been claimed that different representations of the same algebra in the context of the GUP may lead to different physical implications \cite{Bishop22}.
Here, we want to point out a subtle misconception in these arguments that ultimately resolves the ambiguity. In so doing, we shed light on frequently overlooked aspects associated to the GUP that are common sources of misunderstanding and disbelief towards this phenomenological approach to quantum gravity.

To this aim, it is convenient to set the notation and the terminology that will be adopted.
We will address as \emph{physical} the position and momentum operators whose study grants access to the explicit information on the measurement of position and momentum of a given quantum system;
for these operators, we will use the lowercase letters $\hat{x}$ and $\hat{p}$, respectively.
On the other hand, we will address as \emph{canonical} those operators satisfying the Heisenberg algebra;
these will be designated by the uppercase letters $\hat{X}$ and $\hat{P}$ for the position and momentum operators, respectively.
Therefore, while the latter comply with the usual algebra $[\hat{X},\hat{P}] = i \hbar$, the physical operators may not, and in general do not, as in the scenario predicted by the GUP. 

To keep our considerations simple, in line with Refs. \cite{b3,b4,Bishop22,Bishop:2022des}, our reasoning will be carried out in one spatial dimension. This assumption, however, does not harm the validity of our results, which can be promptly generalized to higher dimensions.

\section{A popular example}

Let us consider the case in which the two physical operators $\hat{x}$ and $\hat{p}$ satisfy the most commonly encountered deformed Heisenberg algebra 
\begin{equation}
    [\hat{x},\hat{p}]=i\hbar \left(1+\frac{\beta l_p^2}{\hbar^2}\hat{p}^2\right),\label{qGUP}
\end{equation}
with the Planck length $l_p$ and  the dimensionless model parameter $\beta$. In order to find a viable representation for these observables, we may, for example, express the physical position as a function of the canonical operators while leaving the momentum untouched, that is
\begin{align}\label{posrep}
    \hat{x} =& x(\hat{X},\hat{P}),&
    \hat{p} = \hat{P}.
\end{align}
In this case, up to an irrelevant ordering prescription, we have \cite{Bosso:2021koi}
\begin{align}
    \hat{x} =& \left(1+\frac{\beta l_p^2}{\hbar^2}\hat{P}_1^2\right)\hat{X}_1,&
    \hat{p} =& \hat{P}_1,
    \label{eqns:can_mom_expl_com1}
\end{align}
where the different subscripts for positions and momenta henceforth indicate distinct sets of conjugate variables that are not equivalent.
This labeling turns out to be necessary since the above mapping is not unique, as it does not comprise the only available representation; indeed, there are infinitely many possible transformations which still satisfy the deformed algebra \eqref{qGUP}.
For instance, it is possible to rewrite the physical momentum as a function of the canonical one instead of modifying the position operator, namely
\begin{align}
    \hat{x} =& \hat{X},&
    \hat{p} =& p(\hat{P}).
    \label{eqns:canonical_position}
\end{align}
In phenomenological applications, it is often assumed that the above shape is compatible with \cite{Das:2008kaa}
\begin{align}
    \hat{x} =& \hat{X}_2,&
    \hat{p} =& \hat{P}_2\left(1+\frac{\beta l_p^2}{3\hbar^2}\hat{P}_2^2\right).\label{BCS trafo}
\end{align}
However, it has been recently pointed out \cite{b1,b2,b3,b4,Bishop22,Bishop:2022des} that the representation \eqref{BCS trafo} is consistent with the commutator
\begin{equation}
    [\hat{x},\hat{p}]=i\hbar \left(1+\frac{\beta l_p^2}{\hbar^2}\hat{P}^2\right),\label{BCS com}
\end{equation}
which differs from Eq. \eqref{qGUP} and does not imply a minimal length.
If that were to be the case, the emergence of the minimal length would appear to be representation-dependent.

In quantum theory, it is crucial that predictable physical quantities are independent of the employed Hilbert space basis, \ie, the representation of the algebra of observables.
Correspondingly, this means that a fundamental minimal length derived from a representation-dependent relation could not be observable. A similar outcome would pose a striking problem for the GUP, which should then have to be regarded as an unphysical model; the question thus arises as to how to resolve this apparent ambiguity.

The answer to the conundrum lies within the {perturbative} treatment which is frequently employed when dealing with corrections attributable to the GUP.
Due to the peculiar simplicity of the ensuing eigenvalue problem, the transformation \eqref{BCS trafo} is typically addressed as a \emph{perturbative} approximation in phenomenological applications 
\cite{Das:2008kaa,Pikovski:2011zk,Bawaj:2014cda,Bosso:2017ndq,Bosso:2018ufr,Petruzziello,Bushev:2019zvw}.
The very fact that it leads to the commutator \eqref{BCS com} indicates that it loses its predictability in the \emph{nonperturbative} regime; indeed, since Eq. \eqref{BCS com} is but a truncated expansion of Eq. \eqref{qGUP} in the minimal length factor, it fails when considered at extremely small distances. 

To make the above statement clearer, let us consider a completely general and arbitrary set of {physical} observables satisfying the GUP-deformed Heisenberg algebra 
\begin{equation}
    [\hat{x},\hat{p}]=i\hbar f(\hat{p}),\label{genGUP}
\end{equation}
with a nonsingular, smooth function $f.$ At this point, one can define the canonical operators $\hat{X}$ and $\hat{P}$ such that
\begin{equation}
    \hat{x}=\hat{X}\hspace{1cm}\hat{p}=g(\hat{P}),\label{gen_new_op}
\end{equation}
where $g$ is an invertible, \ie surjective, function. Plugging the transformation \eqref{gen_new_op} into \eqref{genGUP} and using the standard Heisenberg algebra for $\hat{X}$ and $\hat{P}$, we obtain the condition
\begin{equation}
    g'(\hat{P})=f(\hat{p}).\label{FuncCond}
\end{equation}
Note that the left-hand side is a function of $\hat{P},$ whilst the right-hand side depends on $\hat{p}.$ The flaw underlying the transformation \eqref{BCS trafo} consists in setting $\hat{P}=\hat{p}$ in the right-hand side of Eq. \eqref{FuncCond} and requiring the condition to hold true at \emph{all} energy scales, although it is justified only at small momenta, namely when $\beta l_p^2 \braket{\hat{p}^2}\ll 1$. At the nonperturbative level, the relation \eqref{FuncCond} entails
\begin{equation}\label{difeq}
    g'(\hat{P})=f\circ g(\hP),
\end{equation}
thus providing a differential equation which, if complemented with the ordinary quantum mechanical low-energy limit, completely characterizes the function $g.$

For the specific case $f(\hat{p})=(1+\beta l_p^2\hat{p}^2/\hbar^2)$ {contemplated in Eq. \eqref{qGUP}}, the differential problem \eqref{difeq} can be solved analytically, leading to the transformation
\begin{align}
    \hat{x} =& \hat{X}_3,&
    \hat{p} =& \frac{\hbar}{l_p \sqrt{\beta}} \tan \left(\frac{l_p\sqrt{\beta}}{\hbar} \hat{P}_3\right),
    \label{eqns:can_pos_expl_com1}
\end{align}
which for small values of the momentum correctly reduces to Eq. \eqref{BCS trafo}. Before moving on, it is worth stressing that the transformations \eqref{eqns:can_mom_expl_com1} and \eqref{eqns:can_pos_expl_com1} are two different realizations of the operators $\hat{x}$ and $\hat{p}$ in terms of two different pairs of canonical operators $(\hat{X}_1,\hat{P}_1)$ and $(\hat{X}_3,\hat{P}_3)$, that nonetheless correspond to the same deformed function $f(\hat{p})$.

\begin{figure}[t]
    \centering
    \includegraphics[width=\linewidth]{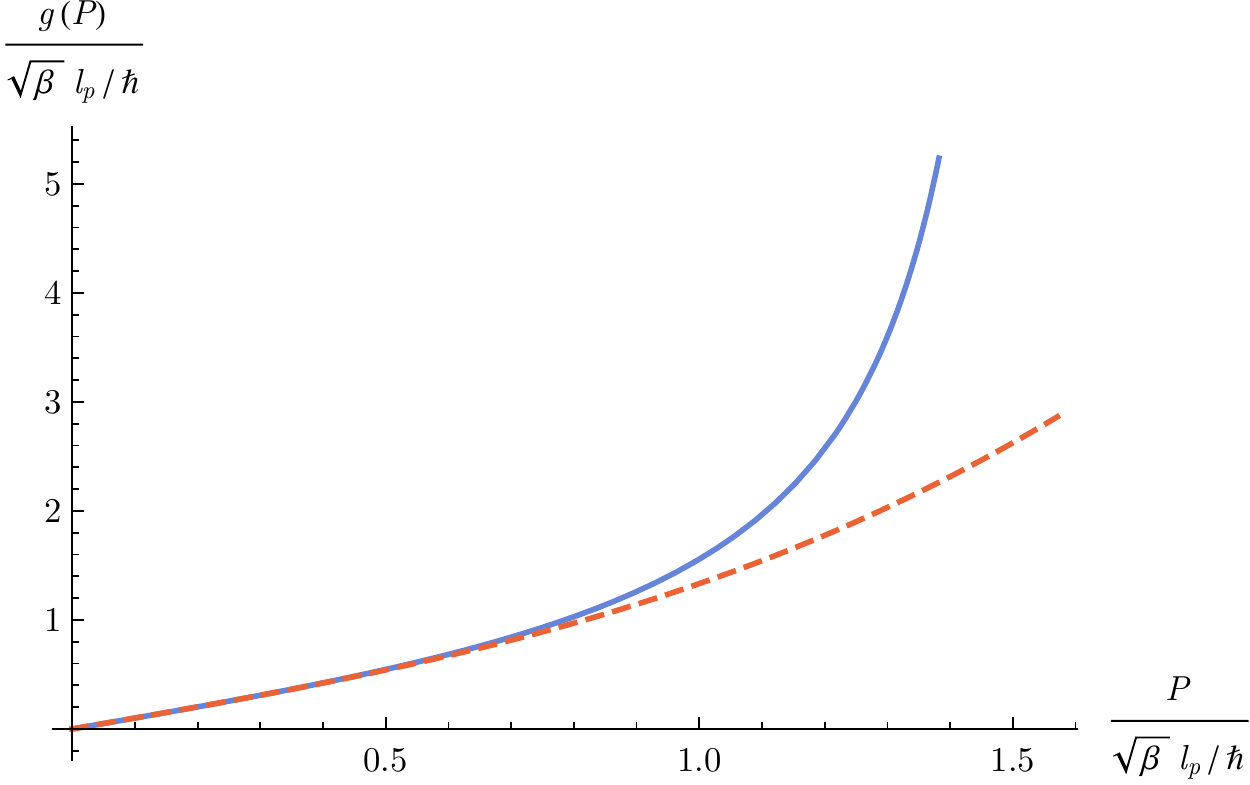}
    \caption{Plot of the physical momentum $p$ given in Eq. \eqref{BCS trafo} (red, dashed) and Eq. \eqref{eqns:can_pos_expl_com1} (blue) as a function of the canonical momentum $P$. Both $g(P)$ and $P$ are measured in units of $\sqrt{\beta}l_p/\hbar.$}
    \label{fig:funccomp}
\end{figure}

For the sake of completeness, a comparison of the result \eqref{eqns:can_pos_expl_com1} with the one expressed by the transformation \eqref{BCS trafo} is plotted in Fig. \ref{fig:funccomp}.
Clearly, the behavior of $\hat{p}$ in \eqref{BCS trafo} significantly deviates from the one in \eqref{eqns:can_pos_expl_com1} for large momenta (\ie, small distances).
Indeed, in the latter scenario the tangent diverges at finite $\hP,$ thus indicating the existence of a minimal length scale.
As already pointed out before, such a high degree of accuracy is not required for phenomenological studies concerning length scales much greater than the minimal length itself; nevertheless, when dealing with the consistency of the GUP  approach, a nonperturbative treatment is mandatory.

\section{The general problem}

To provide a further insight on the topic previously treated, in this section we will tackle the issue raised in Refs.  \cite{Bishop22,b1,b2,b3,b4} from a different perspective.
For the sake of argument, let us introduce on completely general terms a relation of the form
\begin{equation}
    [\hat{x},\hat{p}]=i\hbar F(\hat{P}).
    \label{eqn:CBS_com_gen}
\end{equation}
{While the right-hand side is written as a function of the canonical operator $\hat{P}$, the left-hand side still contains the {physical} quantities $\hat{x}$ and $\hat{p}$.
The operators $\hat{P}$ and $\hat{p}$ are \emph{de facto} distinct, even though they are connected by a functional relation of the kind $\hat{P} = g^{-1}(\hat{p})$, in accordance with Eq. \eqref{gen_new_op}.
It is worth emphasizing one more time that no physical meaning is ascribable to $\hat{P}$, which is only an auxiliary operator introduced to both streamline computations and sharply highlight the deviation from ordinary quantum mechanics. 

Now, when the above functional dependence between $\hat{P}$ and $\hat{p}$ is explicitly embedded in \eqref{eqn:CBS_com_gen}, we derive a relation which formally resembles Eq. \eqref{genGUP} once the identification $f = F \circ g^{-1}$ is accounted for.
By virtue of the Robertson-Schr\"odinger prescription, it is possible to establish a deformed uncertainty relation between the physical operators, that is
\begin{equation}
    \Delta x \Delta p 
    \geq \frac{\hbar}{2} |\langle f(\hat{p}) \rangle|
    = \frac{\hbar}{2} |\langle (F \circ g^{-1})(\hat{p}) \rangle|,
\end{equation}
which, we stress, encodes the entire physical content of the algebra.
The above expression shows that, for a specific function $F$, different choices of the function $g$ (\ie, different representations) not only lead to discrepant values for the minimal length when its presence is predicted, but also encompass completely dissimilar uncertainty relations.
Furthermore, just like a given shape for the function $g$ univocally determines $f$ (and thus the uncertainty relation) once $F$ is given, a specific form of $f$ determines the function $g$ for fixed $F$.
However, whilst $f$ encloses a physical information as it explicitly appears in the uncertainty principle, the functions $F$ and $g$ { by themselves} do not, since any pair of functions satisfying the condition $f = F \circ g^{-1}$  ends up in the same uncertainty relation.
}

To clarify this point, let us consider the commutator \eqref{qGUP} under the assumption contained in Eq.  \eqref{posrep}.
As argued above, this leads to the representation provided in Eq. \eqref{eqns:can_mom_expl_com1} which is compatible with the results in \cite{Kempf:1994su,Bosso:2020aqm} but differs from \eqref{BCS trafo}.
We have thus found two possible alternatives for the function $g$ with the same physical implications.
These alternatives are paired with a specific shape for $F$, that is
\begin{align}
    &\text{Eq. \eqref{eqns:can_mom_expl_com1}}  \Longrightarrow &
    F(\hat{P}) &= 1 + \frac{\beta l_p^2}{\hbar^2} \hat{P}^2,\\
    &\text{Eq. \eqref{BCS trafo}} \Longrightarrow &
    F(\hat{P}) &= 1+\frac{\beta l_p^2}{\hbar^2}\hat{P}\left(1-\frac{\beta l_p^2}{3\hbar^2}\hat{P}^2\right)^2.
\end{align}
%
As a further example, let us consider the case $f(\hat{p}) = 1$, corresponding to ordinary quantum mechanics.
This physical scenario (the Heisenberg principle) can be described in terms of any invertible function $F$.
Indeed, by choosing $g = F$ we obtain $f = F \circ g^{-1} = 1$.
This peculiar case manifestly indicates that the function $F$ has no physical attributes, and therefore the characterization of a model based on such a function is not correct; as argued before, it is evident that the same consideration also holds for $g$.
In particular, it is worth highlighting once more that the equivalent choices \eqref{eqns:can_mom_expl_com1} and \eqref{eqns:can_pos_expl_com1} referred to the commutator \eqref{qGUP} are indeed examples of a wide class of viable alternatives, and hence carry no physical meaning.
The true physical significance lies in the specific relation between the physical observables $\hat{x}$ and $\hat{p}$, that is, the commutator.


\section{Concluding remarks}

In this letter, we have clarified that the physical content on quantum mechanics (both the ordinary and the deformed versions) is encoded within the commutators of \emph{physical} observables, with which the Hamiltonian (responsible for the dynamics) is built. According to the very axioms of quantum mechanics, the way we decide to represent the algebra of observables when constructing a Hilbert space is irrelevant. In turn, it is important to distinguish between modifications of the underlying algebra for physical operators and changes of a basis in Hilbert space instead of blurring the distinction between physics and representations. 

For the problem at hand,
we have remarked how the ambiguity recently raised in literature concerning the GUP \cite{b3,b4,Bishop22,Bishop:2022des} is the result of an improper treatment of the perturbative expansion in the minimal length. Despite the fact that this approximation is allowed in phenomenological studies due to the huge gap between the experimentally reachable scales and the regime where quantum gravitational effects become relevant, a consistency check of the GUP approach cannot but be performed by means of a nonperturbative analysis.

In light of this finding, we thus conclude that the characteristic implications of the GUP (such as the appearance of a minimal observable length) are physical and cannot be altered by different representations of the {physical} operators obeying the deformed algebra.

\medskip
\section*{Acknowledgements}
\vspace{-1em}

 The authors would like to thank D. Singleton and M. Bishop for helpful discussions.
 F.W. is thankful to the quantum gravity group at CP$^3$ origins, University of Southern Denmark, for its kind hospitality during his research visit, and was supported by the Polish National Research and Development Center (NCBR) project ''UNIWERSYTET 2.0. --  STREFA KARIERY'', POWR.03.05.00-00-Z064/17-00 (2018-2022). The authors acknowledge networking support by the COST Action CA18108.


\end{document}